
\input harvmac
\def\rhob{{\rho\kern-0.465em \rho}}

\def\o{\over}

\def\ontopss#1#2#3#4{\raise#4ex \hbox{#1}\mkern-#3mu {#2}}
\def\half{{1 \over 2}}
\def\noN{{n \over N}}
\def\ul{{\bigl\lfloor{N-1 \over 2}\bigr\rfloor}}
\def\ula{{\bigl\lfloor{a-1 \over 2}\bigr\rfloor}}
\def\ulNa{{\bigl\lfloor{N-a-1 \over 2}\bigr\rfloor}}
\def\lla{{\bigl\lfloor{a+2 \over 2}\bigr\rfloor}}
\def\om{\omega}

\setbox\strutbox=\hbox{\vrule height12pt depth5pt width0pt}

\def\strut{\relax\ifmmode\copy\strutbox\else\unhcopy\strutbox\fi}


\nref\rbaxa{R.J. Baxter, in Fundamental Studies in Statistical
Mechanics V, E.G.D. Cohen ed., North Holland, (1980) 109.}

\nref\rbaxb{R.J. Baxter, Exactly Solved Models in Statistical
Mechanics, Academic Press (1982).}
\nref\rs{N. Seiberg, Phys. Lett. B 318 (1993) 469 and Phys. Rev. D 49 6857.}

\nref\rsw{N. Seiberg and E. Witten, Nucl. Phys. B 426 (1994) 12 and B
431 (1994) 482.}
\nref\rampty{H. Au-Yang, B.M. McCoy, J.H.H. Perk, S. Tang, and
M--L. Yan, Phys. Lett. A123 (1987) 219.}

\nref\rbpa{R.J. Baxter, J.H.H. Perk and H. Au-Yang, Phys. Lett. A
128 (1988) 138.}
\nref\rap{H. Au-Yang and J.H.H. Perk, Advanced Studies in Pure Math. 19
(189) 57.}
\nref\rampc{G. Albertini, B.M. McCoy and J.H.H. Perk, Advanced Studies
in Pure Mathematics 19 (1989) 1.}
\nref\rbaxc{R.J. Baxter, Phys. Lett. A 146 (1990) 110; Free Energy of the
Chiral Potts Model in the Scaling Region, Canberra preprint (1995)}
\nref\rampa{G. Albertini, B.M. McCoy, J.H.H. Perk and S. Tang,
Nucl. Phys. B314 (1989) 159.}
\nref\rampb{G. Albertini, B.M. McCoy and J.H.H. Perk, Phys. Lett. A135
(1989) 159.}
\nref\rpert{J.J. Sakurai, Modern Quantum Mechanics, Addison-Wesley (1985).}
\nref\rhanhon{N.S. Han and A. Honecker, Jour. Phys. A 27 (1994) 9.}
\nref\rhona{A. Honecker, Quantum Spin Models and Extended Conformal
Algebra, HEPTH 9503104.}
\nref\rvgh{ A. Honecker and G. von Gehlen, Nucl. Phys. B435 (1985) 505.}
\nref\rhonb{A. Honecker, A Perturbative Approach to the Chiral Potts
Model, HEPTH 9409122.}
\nref\rag{ I.G. Enting and A.J. Guttmann, The Ising model, self
avoiding walks and solvability, Lattice '95 Proceedings, Nucl. Phys. B
suppl (submitted).}
\nref\rwmtb{T.T. Wu, B.M. McCoy, C.A Tracy and E. Barouch, Phys. Rev,
B13 (1976) 316.}
\nref\rtracy{C.A. Tracy, in Nonlinear Equations in Physics and
Mathematics, A.O. Barut (ed.), D. Reidel Publ. Co., Dorttrecht,
Holland, (1978) 221.}
\nref\rbaxe{R.J. Baxter, Phil. Trans. Roy. Soc. Lond. 289 (1978) 315.}
\nref\rbaxf{R.J. Baxter, J. Stat. Phys. 63 (1991) 433; 70 (1993) 535.}
\nref\rhkd{S. Howes, L.P. Kadanoff and M. den Nijs, Nucl. Phys. B215
[FS7] (1983) 169.}
\nref\rmehta{ A. Gervois and M.L. Mehta, Jour. Math. Phys. (in press).}
\nref\rgaz{S. Gasiorowicz, Quantum Physics, Wiley (1974).}
\nref\rbate{A. Erdelyi, Higher Transcendental Functions, vol.~1, McGraw-Hill
(1953).}
\nref\rons{L. Onsager, Phys. Rev. 65 (1944) 117.}
\nref\rrade{H. Rademacher and E. Grosswald, Dedekind Sums, Carus Mathematical
Monograph No.~16, Mathematical Association of America (1972).}
\nref\ras{M. Abramowitz and I. Stegun,  Handbook of Mathematical Functions
with Formulas, Graphs, and Mathematical Tables, U.~S.~Government
Printing Office (1972).}
\Title{\vbox{\baselineskip12pt\hbox{ITPSB 95-14}
  \hbox{HEP-TH 9509004}}}
  {\vbox{\centerline{Analyticity and Integrability}
  \centerline{in the}
  \centerline{Chiral Potts Model}}}

  \centerline{ Barry~M.~McCoy~\foot{mccoy@max.physics.sunysb.edu}~and
William~P.~Orrick~\foot{worrick@insti.physics.sunysb.edu}}

  \bigskip\centerline{\it Institute for Theoretical Physics}
  \centerline{\it State University of New York}
  \centerline{\it  Stony Brook,  NY 11794-3840}

\Date{\hfill 9/95}

  \eject

  \centerline{\bf Abstract}
We study the perturbation theory for the general non-integrable chiral
Potts model depending on two chiral angles and a strength parameter
 and show how the analyticity of the ground state energy and
correlation functions
 dramatically increases when the angles and the strength parameter
satisfy the integrability condition.  We further specialize to the
superintegrable case and verify that a sum rule is obeyed.

  \newsec{Introduction}
 The connection between integrability and analyticity has been
recognized and exploited for many years in the theory of solvable
models in statistical mechanics and in particular in the method of
solution based on the inversion relation~\rbaxa-\rbaxb~where assumptions on
analyticity are a vital ingredient of the
solution. Conversely, whenever an integrable model can be solved by
other means the analyticity demanded by the inversion relation is
always found to exist.

On the other hand these analyticity statements have never been made
particularly precise and, indeed there are recent computations in
certain four dimensional field theories~\rs-\rsw~where results may be
obtained because functions also have dramatically expanded analyticity
properties over what is generically expected but no analogue of the
two dimensional integrability conditions is known. Consequently it is
of interest to sharpen our understanding of the connection between
integrability and analyticity.

In this note we examine these questions for the $N$ state chiral Potts
spin chain defined by
\eqn\cp{H_{{\rm CP}}=-\sum_{j=1}^L\sum_{n=1}^{N-1}\left({e^{i(2n-N)\phi/N}\over
\sin(\pi n/N)}(Z_jZ^{\dagger}_{j+1})^n~\
+k{e^{i(2n-N){\bar
\phi}/N}\over \sin(\pi n/N)}X_j^n~\right).}
The matrices $X_j,Z_j$ are defined by
\eqn\exj{X_j=I_N\otimes\cdots\underbrace{X}_{{\rm
site}~j}\cdots\otimes I_N~,\qquad
Z_j=I_N\otimes\cdots\underbrace{Z}_{{\rm
site}~j}\cdots\otimes I_N~,}
where $I_N$ is the $N\times N$ identity matrix, the elements of the $N\times
N$ matrices $X$ and $Z$ are
\eqn\exel{X_{l,m}=\delta_{l,m+1}~({\rm mod}~N)~,\qquad
Z_{l,m}=\delta_{l,m}\omega^{l-1}}
and $\omega=e^{2\pi i/N}.$ When the angles $\phi,~{\bar \phi},$ and
the strength parameter $k$ satisfy the condition
\eqn\int{\cos \phi=k\cos {\bar \phi}}
this spin chain Hamiltonian is derivable from the $N$ state chiral
Potts model whose transfer matrix $T_{p,q}$ satisfies the integrability
condition of commuting transfer matrices~\rampty
\eqn\com{\left[ T_{p,q},T_{p,q'}\right]=0}
where $p$ and $q$ specify points on the Riemann surface given by
the intersection of two Fermat cylinders~\rbpa-\rap
\eqn\curve{a^N+kb^N=k'd^N,~~~ka^N+b^N=k'c^N}
with $k^2+k'^2=1.$ The Hamiltonian \cp~is obtained from $T_{p,q}$ in
the limit $q\rightarrow p$ as~\rampc
\eqn\hlim{T_{p,q}=I+u(H_{CP}+{\rm const})+O(u^2)}
where $u$ measures the deviation of $q$ from $p$ and
\eqn\angles{e^{{2i\phi\over N}}=\omega^{1/2}{a_p c_p\over b_p
d_p},~~~\
e^{{2i{\bar \phi}\over N}}=\omega^{1/2}{a_p d_p\over b_p c_p}.}

The free energy of the statistical mechanical system and the ground
state energy of the spin chain have been computed in the general
integrable case~\rbaxc~and in the special superintegrable case of $\phi={\bar
\phi}=\pi/2$ where there is a dramatic simplification~\rampc\rampa-\rampb.
Our study here
is to relate the general non-integrable case to the integrable and
superintegrable cases.

For the non-integrable case the only analytic tool
available is perturbation theory in the variable $k$. This expansion
can be done both for $k\sim 0$ and $k\sim \infty.$  In this paper we
study only the first case and consider the expansion for the $N$ state
spin chain of the ground state energy and correlation functions.
We observe that as the order increases, new singularities in
the $e^{i\phi}$ plane continually arise and we expect that the
number of singularities will become infinite as the order goes to
infinity.  The integrability condition~\int,~however, involves $k$
and causes extensive cancelations among orders.  We find that all but a finite
number of singularities, which are already present in the lowest orders,
vanish resulting in a dramatic increase in analyticity.  This increase
in analyticity is intimately related to the integrability of the model.

In Section~2~we present the perturbation expansion of the ground
state energy for general chiral angles $\phi$ and $\bar\phi$
for $N=3$--$6$ and then
exhibit the increase in analyticity that occurs when the
integrability condition~\int~is imposed.  In Section~3~the same
procedure is applied to the correlation functions for $N=3$.  In
addition, the perturbation expansion for the superintegrable
correlation functions is given for arbitrary $N$.  In Section~4~a
sum rule for the nearest neighbor superintegrable correlation
functions is derived and used to verify the expansion in Section~3.
Also, a Fuchsian equation for the superintegrable ground state
energy is derived.  In Section~5~we conclude with some questions
for further study and some general remarks on the connection
between integrability and analyticity.

\newsec{Perturbation Expansion of the Ground State Energy}

In this section we consider the expansion for the $N$ state
chiral Potts spin chain with general chiral angles, $\phi$ and
$\bar\phi$, of the ground state energy
\eqn\pert{e^{(N)}=\lim_{L\rightarrow \infty}E^{(N)}_{GS}/L=\
\sum_{n=0}^{\infty}k^n e^{(N)}_n.}
We have implemented Rayleigh-Schroedinger perturbation theory~\rpert~on the
computer using a program written in C, Mathematica, Maple and Form and
studied the following cases:

$N$=3 to order 9~~~$N$=4 to order 8~~~$N$=5 to order 5~~~$N$=6 to order 6.

Our  results for these 4 cases are given in Tables~1--4 where we use the
following notation
\eqn\cdefin{\eqalign{C_3&=\cos(\phi/3),~~~C_4={\sqrt 2}\cos(\phi/2)\cr
C_5&={\cos(\phi/5)\over\sin(\pi/5)}=
{\sqrt 2}{\sqrt{ 1+5^{-1/2}}}\cos(\phi/5),~~~C_6={\sqrt 3}\cos(\phi/3).}}
In the case of
$N=3$ there have been previous perturbation studies both for
small~\rhanhon-\rhona~and large~\rhona-\rhonb~$k$ and Table~1 agrees with
the small $k$ results of~\rhona~which were obtained to order 7.
Many of the details
of the perturbation expansion are discussed in these papers.

These  tables demonstrate an important feature of the generic
perturbation theory; namely that as the order of perturbation theory
increases so do the number of places in the $e^{i\phi}$ plane where
the perturbation expansion develops singularities. Thus, for example,
we see from Table~1 that $e^{(3)}_2$ and $e^{(3)}_3$ are singular only when
$C_3=0$ but that $e^{(3)}_4$, $e^{(3)}_5$, $e^{(3)}_6$, and
$e^{(3)}_7$and  have
additional singularities at $C_3=\pm 1/2$ and $e^{(3)}_8$
and $e^{(3)}_9$ have further
singularities at $28C_3^2-3=0.$ A similar increase in number of
singularities is seen in Table~2. Indeed this property
of the  number of singularities increasing with the order has been
explicitly seen in a variety of previous perturbation computations
including the Ising model in a magnetic field~\rbaxa~and the
zero field Ising magnetic susceptibility~\rag. In the latter case is it well
understood that  the number of singularities is connected to the
number of quasiparticles that contribute to the given order of
perturbation theory~\rwmtb-\rtracy~and that as the order of perturbation goes
to infinity an infinite number of singularities will appear. It is
this ever increasing number of singularities in the perturbation
theory which in the case of the Ising model in a magnetic field
restricts the analyticity to such an extent~\rbaxa~that a closed form
solution has never been found.

However there is a dramatic change in this pattern of singularities
when the parameters $\phi,~{\bar \phi}$, and $k$ are related by the
integrability condition~\int. This is possible because this
integrability condition~\int~involves the perturbation parameter $k$
and thus allows different orders of perturbation theory to be
combined.

To appreciate the cancelations between the orders of
perturbation theory that result we consider $N=3$ in detail.
It is easy to see
when~\int~holds that if we use
\eqn\relation{{\cos \phi\over \cos(\phi
/3)}=e^{2i\phi/3}-1+e^{-2i\phi/3}}
then $e^{(3)}_2$ and $e^{(3)}_3$ are combined as
\eqn\spt{k^2e^{(3)}_2+k^3e^{(3)}_3=-k^2{3+4C_3^2\over 9 {\sqrt 3} C_3}}
which is the $k^2$ term on the integrable manifold as given in Table~5.

A more important cancelation, however, takes place between the terms
$e^{(3)}_4$, $e^{(3)}_5$ and $e^{(3)}_6$ to give the $k^4$ term
in Table~5. Here the
terms in $\cos{\bar \phi}$ in the $k^5$ terms and the terms in
$\cos^2{\bar \phi}$ in the $k^6$ term which have the denominators
$\pm 1+2 C_3$ combine with the terms in the $k^4$ term with the same
denominators to cancel the denominators out completely.  In addition
the terms with $C_3^{-4}$ in $e^{(3)}_5$ and with $C_3^{-5}$ in $e^{(3)}_6$ are
reduced to $C_3^{-3}$ by use of~\relation~and thus the complete $k^4$ term in
Table~5 is obtained from three terms in Table~1. Similarly, the
terms in $e^{(3)}_6$ which have no factor of $\cos{\bar \phi}$  combine with
the terms in $e^{(3)}_7$ with $\cos{\bar\phi}$, the terms in $e_8$ with
$\cos^2{\bar\phi}$ and the terms in $e^{(3)}_9$ with $\cos^3{\bar\phi}$ to
cancel all denominators containing powers of $\pm1+2C_3.$ Thus we
obtain the term of order $k^6$ in Table~3.

For $N=4$ we find from Table~2 that the terms of orders 2, 3 and 4 are
needed to get the $k^2$ term on the integrable manifold and that to
get the $k^4$ term in the integrable case the terms in the generic
case to order 8  are needed. In
general to get the order $k^{2q}$ term in the integrable case the
terms to order $Nq$ of the generic case are needed.  To show this
we first note that $e^{(N)}$ is an even function of $\bar \phi$.
Secondly, the order $r$ term in the generic case is a sum of terms
of the form
\eqn\perttm{C_{{\bf n},{\bf j}}
(\phi) k^r \exp \left({i {\bar\phi} \o N}\left(2
\sum_{l=1}^r n_l -Nr\right)\right) _{0\!\!\!\!}\left\langle 0 \left |
\prod_{l=1}^r X_{j_l}^{n_l}\right | 0 \right\rangle_{\!\!0}}
where $|0\rangle_0$ is the ground state when $k=0$.
The product
clearly vanishes if $\sum_{l=1}^r n_l$ is not an integer multiple of $N$.
Thus the $k^r$ term can be written
\eqn\perttma{e^{(N)}= \sum_{r=0}^\infty k^r \sum_p C_p(\phi) \cos(p {\bar\phi})
= \sum_{r=0}^\infty k^r \sum_p {C'}_p(\phi) \cos^p{\bar\phi}}
where the sum over $p$ is taken over $p$ even if $r$ is even and $p$ odd
if $r$ is odd.  When the integrability condition~\int~is applied we obtain
\eqn\perttmb{e^{(N)}=\sum_{r=0}^\infty \sum_p k^{r-p} {C'}_p(\phi) \cos^p\phi}
which contains only even powers of $k$.
Since the maximum allowed value for the $n_l$ in~\perttm~is $N-1$, the
maximum allowed value of $p$ for a given $r$ is $p_{\rm max}(r)=
2\left\lfloor{r (N-1)\o N}\right\rfloor - r$.  The only terms in the
generic case contributing to the $k^{2q}$ term in the integrable case
are those terms for which $r-p_{\rm max}(r)\le 2q$ which is
equivalent to the condition $-\left\lfloor -{r\o N}\right\rfloor  \le q$
or $r \le q N$ as desired.

The results of the
specialization to the integrable manifold are given in Table~5 for the
cases $N=3$--$6$.  Thus we see in the cases considered that all
singularities have canceled out except certain of the original
singularities in $e^{(N)}_2.$ Consequently the analyticity of the ground state
energy has increased to the maximum extent possible. This remarkable
extension in analyticity is a
consequence of the defining commutation relation of integrability~\com~
and the relation~\angles~which the chiral angles have with the
Riemann surface~\curve.

These integrable results are now easily specialized to the
superintegrable case $\phi=\pi/2$ and we obtain
\eqn\sigs{
\eqalign{
e^{(3)}& =-2-{4 \over {3^2}} k^2 -{8 \over {3^5}} k^4 - {{20} \over {3^7}} k^6
-O(k^8) \cr
\noalign{\smallskip}
e^{(4)}& = -3-{5 \over {2^3}} k^2 -{{25} \over {2^9}} k^4 - O(k^6)\cr
\noalign{\smallskip}
e^{(5)}& = -4-{4 \over 5} k^2 - O(k^4)\cr
\noalign{\smallskip}
e^{(6)}&=-5-{35\over 36}k^2-O(k^4)\cr}}
which agrees with the previous computation
of~\rampa. We note that in principle we should compare the
integrable results of Table~5 with the exact result of~\rbaxc~but no
systematic expansion of the exact  result seems to be in the literature.

\newsec{Perturbation Expansion of the Correlation Functions}

The cancelation of singularities and resulting expansion of
analyticity seen in the ground state is, of, course, expected to occur
in all the correlation functions as well. However, the presentation of
the results of the perturbation theory in the generic case is more
cumbersome and consequently we will restrict ourselves here to
illustrating the phenomena for the case $N=3.$

The perturbation expansion of the correlation function
$\langle Z_0 Z^{\dagger}_r \rangle$ for generic values of $\phi$ and
$\bar \phi$ was studied in~\rhanhon-\rhona~to fourth
order. In order to exhibit the desired cancelations of the poles
at $C_3=\pm 1/2,$ which occurred in the ground state energy by combining
the fourth, fifth and sixth orders, we must extend the computation of
the correlation to sixth order also. This we have done and the
results are given in Table~6. The method of perturbation is discussed
in some detail in~\rhanhon-\rhona. The only additional remark we wish
to make is that in contrast to more standard perturbation expansions
we have been unable to find an efficient way to perform the
subtraction of the disconnected diagrams to obtain a general formalism
that deals only with order one objects in the $L\rightarrow \infty$
limit and are forced to perform an explicit subtraction of $L$
dependent terms. This seems to be a consequence of doing perturbation
theory on a lattice with a variable that satisfies a cyclicity condition
$X_j^N=1$ and, because of the connection of the cyclicity condition
with the fractional statistics of the model, it deserves further study.

To study the specialization of the generic result of Table~6 to the
integrable case~\int~we consider first the long range order
term $M^2=\lim_{r\rightarrow \infty}\langle Z_0 Z_r^{\dagger}\rangle$
which in the
perturbation expansion is the coefficient of $1-\delta_{n,0}.$ From
the second and third order terms it is easy to see that the
cancelation of poles is more complete than in the ground state energy
and that to second order on the integrable manifold we have the
simplification to $-2k^2/9$ where the dependence on the chiral angle
$\phi$ has completely vanished and thus the order parameter on the
integrable manifold is identical with the order parameter on the
superintegrable special case. This complete cancelation occurs in
all orders and is a well known consequence~\rbaxe-\rbaxf~of the integrability
commutation relation~\com. These perturbation computations for the
long range order have been
carried to high order in $k$ for $N=3$ in~\rbaxf-\rhkd~and have
been extended to
arbitrary $N$ in~\rampa. These expansions have
lead to the remarkably
simple conjecture for the general case that
\eqn\conj{M_n^2=\lim_{r\rightarrow \infty}\langle (Z_0
Z_r^{\dagger})^n\rangle=(1-k^2)^{n(N-n)/N^2}.}

To see the first dependence on the variable $\phi$ in the correlation
function on the integrable manifold we must go to fourth order
 which is obtained from the fourth, fifth and sixth orders of of the
generic result of Table~6.  (Note that to obtain the $k^{2q}$ term
in the integrable case we need terms up to $k^{Nq}$ in the generic
case just as we did for the ground state energy.)
In fourth order the only deviation of $\langle Z_0 Z_r^{\dagger}\rangle$
from $M_1^2$ occurs for $r=0,1$ and
since the case $r=0$ is trivial we consider only  $r=1$ and find that
the poles at $C_3=\pm 1/2$ cancel. Thus
the result is obtained on the
integrable manifold~\int~of
\eqn\intcorr{\langle Z_0 Z_1^{\dagger}\rangle-M_1^2=k^4\bigl({7\over
162}-{1\over 108 \cos^2\phi/3}+
{i\tan{\phi/3}\over 54 }\bigr)+ O(k^6).}

This result is to be compared with the fourth order result on the
superintegrable manifold which may be obtained for arbitrary $N$.
{}From a straightforward perturbation expansion we may reduce the
correlation to a set of trigonometric sums
\eqn\supcorra{\eqalign{&\langle (Z_0 Z_1^{\dagger})^n\rangle
=1-{n \over N} \left(1-{n \over N}\right) k^2 \cr
&+\biggl\{{1 \over 16N^4}\sum_{j=1}^{N-1} \sum_{k=1}^{N-1} {2(1-\omega^{nj})
+2(1-\omega^{-nk})-(1-\omega^{n(j-k)}) \over \sin^2 \pi j/N \sin^2
\pi k/N} \cr
&-{1 \over 8N^4}\sum_{p=1}^{N-1}\left( 1-\cos {2 \pi np \over N}\right)
\sum_{j=1 \atop j\neq p}^{N-1} \sum_{k=1 \atop k\neq p}^{N-1}
\left[ \sin {\pi j \over N} \sin {\pi (p-j) \over N}
\sin {\pi k \over N}\sin {\pi (p-k) \over N}\right]^{-1} \cr
&-{1 \over 4N^4}\sum_{p=1}^{N-1} {(1-\cos 2 \pi np/N) \over \sin^4
\pi p/N} \cr
&+{1 \over 4N^4}\sum_{p=1}^{N-1}\left( 1-\cos {2 \pi np \over N}\right)
\sum_{j=1 \atop j\neq p}^{N-1} \sum_{k=1 \atop k\neq j}^{N-1}
\left[ \sin {\pi (p-j) \over N} \sin {\pi k \over N}
\sin {\pi (j-k) \over N}\sin {\pi p \over N}\right]^{-1} \cr
&-{3 \over 16N^4} \sum_{j=1}^{N-1} \sum_{k=1}^j {(1-\omega^{n(j-k)})
+(1-\omega^{-nj})+(1-\omega^{nk}) \over
\sin^2 \pi j/N \sin^2 \pi k/N} \biggr\} k^4 + O(k^6).}}
The trigonometric sums have been evaluated in ref.~\rmehta~and thus
the final result is
\eqn\supcorr{\eqalign{\langle (Z_0 Z_1^{\dagger})^n\rangle-M_n^2
=&{\biggl(}{1\over 2} \left[ {n\over N}\left(1-{n\over N}\right)\right]^2+
{1\over 4N^2}{n\over N}\left(1-{n\over N}\right)\cr
&-i {3\over4 N^4}\sum_{p=1}^{N-1}f(N,n,p)\tan{\pi p\over 2N}{\biggr)} k^4+
O(k^6)}}
where
\eqn\fde{f(N,n,p)=\cases{n^2(N-2p)& if $1\leq n\leq p\leq N/2$\cr
f(N,p,n)& in general \cr
-f(N,N-n,p)& in general. \cr}}
Thus if we set $N=3$ in~\supcorr~we obtain~\intcorr~with $\phi={\pi \o 2}$.

\newsec{Sum Rule for Superintegrable Correlation Functions}

The superintegrable nearest neighbor correlation functions
$\langle(Z_0 Z_1^{\dagger})^n \rangle$ obey a sum rule which provides
a check of the perturbation expansion~\supcorr.  Here we derive several
forms for the sum rule.

We start by applying the Feynman-Hellmann formula~\rgaz
\eqn\feynman{{\partial e^{(N)} \over \partial k}={1 \over L}
\left\langle 0 \biggl| {\partial H \over \partial k} \biggr|0 \right\rangle,}
to~\cp~to obtain
\eqn\suma{\sum_{n=1}^{N-1} \langle 0| \alpha_n (Z_0 Z_1^\dagger)^n |0\rangle
=-e^{(N)} + k {\partial e^{(N)} \over \partial k}}
where
\eqn\alphadef{\alpha_n={e^{i(2n-N)\phi/N}\over \sin(\pi n/N)}}
and the ground state energy per site of the $N$-state superintegrable chiral
Potts chain is~\rampc
\eqn\sigs{e^{(N)}(k)=-(1+k) \sum_{l=1}^{N-1} F\left(-{1\over2},{l\over N};
1;{{4k}\over{(1+k)^2}}\right).}
With the aid of~\rbate~(p.~102 (21))
\eqn\divhyp{F'(a,b;c;z)={a \over z} [F(a+1,b;c;z)-F(a,b;c;z)],}
\suma~and~\sigs~yield a sum rule
\eqn\sumb{\eqalign{\sum_{n=1}^{N-1} \langle 0| \alpha_n (Z_0 Z_1^\dagger)^n
|0\rangle
=&\sum_{n=1}^{N-1} \biggl[ \half (1-k) F\left(\half,{n \over N};1;{4k\over
(1+k)^2}\right) \cr
&+\half (1+k) F\left( -\half,{n\over N};1;{4k\over(1+k)^2}
\right)\biggr].}}

We may reexpress the sum rule in terms of the functions
$F({n\over N},{n\over N};1,k^2)$ and $F(-{n\over N},{n\over N};1,k^2)$
by using the two identities
\eqn\hypida{\eqalign{F\left(\half,b;1;{4k\over(1+k)^2}\right) =& (1+k)^{2b}
F(b,b;1;k^2) \cr
F\left(-\half,b;1;{4k\over(1+k)^2}\right) =& (1+k)^{2b-2}
\left\{4(1-b)F(b,-b;1;k^2) \right. \cr
&+\left.\left[(4b-3)-k \right](1-k)F(b,b;1;k^2)\right\}.}}
The first of these is just
eq.~(36), p.~113 of~\rbate~with $a=b$.  The second can be
obtained from
\eqn\hypidc{\eqalign{F\left(-\half,b;1;{4k\over(1+k)^2}\right)=&
(2b-1)\left(1-{4k\over(1+k)^2}\right)F\left(\half,b;1;{4k\over(1+k)^2}\right)
\cr
&+2(1-b)F\left(\half,b-1;1;{4k\over(1+k)^2}\right)}}
(eq.~(37), p.~103 of~\rbate~with $a=\half$, $c=1$ and $z=4k/(1+k)^2$) by
using the first identity and eq.~(33), p.~103 of~\rbate~with
$a=b-1$, $c=1$ and $z=k^2$.
The sum rule now becomes
\eqn\sumc{\eqalign{\sum_{n=1}^{N-1} \langle 0| \alpha_n (Z_0 Z_1^\dagger)^n
|0\rangle
=&\sum_{n=1}^{N-1}\biggl[\left({2n\over N}-1\right)(1-k^2)(1-k)^{{2n\over N}-2}
F\left(\noN,\noN;1;k^2\right) \cr
&+2\left(1-\noN\right)(1+k)^{{2n\over N}-1}
F\left(\noN-1,\noN;1;k^2\right)\biggr].}}

The two sides of the sum rule are even functions of $k$.  This is made
manifest by adding~\sumc~with
$n\to N-n$ to itself and using eq.~(2), p.~105 and eq.~(31), p.~103
of~\rbate~to obtain
\eqn\sumrule{\eqalign{
&\sum_{n=1}^{N-1} {1 \over {\sin \pi n/N}} \langle 0 | e^{i
{\pi (2n-N)\over 2N}} (Z_0 Z_1^{\dagger})^n + e^{-i {\pi (2n-N) \over 2N}}
(Z_0 Z_1^{\dagger})^{N-n} |0 \rangle \cr
&~~~= \sum_{n=1}^{N-1} \left\{ 2\left( 1-{n \over N}\right)
\left[(1+k)^{{2n \over N}-1}
+(1-k)^{{2n \over N}-1} \right]
F\left( {n \over N}-1,{n \over N};1;k^2 \right) \right. \cr
&~~~~~~\left. + \left( {2n \over N}-1\right) (1-k^2)
\left[ (1+k)^{{2n \over N}-2} + (1-k)^{{2n \over N}-2}\right]
F\left( {n \over N},{n \over N};1;k^2\right) \right\}.}}

Expanding the right hand side of~\sumrule~to fourth order and
using~\supcorr~in the left hand side
the $k^4$ term in the sum rule gives
\eqn\trigida{\sum_{j=1}^{N-1} \sum_{k=1}^{N-1} f(N,j,k) \cot {\pi j \over N}
\cot {\pi k \over N} = {N(N^2-1)(N^2-4) \over 90}.}
This identity is a special case of the more general identity, which is
independently proven in Appendix~A
\eqn\trigidb{\eqalign{S(N)=&\sum_{j=1}^{N-1} \sum_{k=1}^{N-1} f_Q(N,j,k)
\cot {\pi j \over N} \cot {\pi k \over N} \cr
=& {2(N-1)(N-2) \over 3} Q(N,0) -
4 \sum_{j=1}^\ul (N-2j) Q \left( N,{j \over N} \right)}}
where
\eqn\fgen{f_Q(N,j,k)=\cases{P(N,j,k)& if $1\leq j\leq k\leq N/2$\cr
f_Q(N,k,j)& in general \cr
-f_Q(N,N-j,k)& in general \cr}}
and
\eqn\pdef{P(N,j,k) = Q \left( N,{k+j \over N} \right) +
Q \left( N,{k-j \over N} \right) -
2 Q \left( N,{k \over N} \right) }
for any function $Q(N,x)$ satisfying
\eqn\qsymm{Q(N,1-x)=-Q(N,x).}
We note
that~\trigidb~is a member of a broader class of identities, described
in Appendix~B.

We conclude this section by deriving an order $2N-2$ Fuchsian differential
equation for the superintegrable ground state energy~\sigs~with three
singularities located at 0, 1 and $\infty$.  Equation~\sigs~is expressed
in terms of hypergeometric functions which are solutions to the
second order Fuchsian equation with singularities as above
\eqn\hyp{D_{(a,b;c;z)}F(a,b;c;z) = 0}
where
\eqn\hypop{D_{(a,b;c;z)} = z(1-z){d^2 \o d^2z}+\left[c-(a+b+1)z\right]
{d \o dz}-ab.}
One may readily verify that this differential operator satisfies
\eqn\opid{D_{(a+1,b_1;c;z)}D_{(a,b_2;c;z)}=D_{(a+1,b_2;c;z)}D_{(a,b_1;c;z)}.}
We also define the differential operator
\eqn\hypopp{D'_{(a,b;c;k)}=(1+k)D_{\left(a,b;c;{4k \o (1+k)^2}\right)}
(1+k)^{-1}}
which has solution $(1+k)F\left(a,b;c;{4k\o (1+k)^2}\right)$ and which
satisfies an equation precisely analogous to~\opid. Now let
\eqn\dprod{D_N=D'_{\left({2N-5\o 2},{N-1\o N};1;k\right)}
D'_{\left({2N-7\o 2},{N-2\o N};1;k\right)}\ldots
D'_{\left(-{1\o 2},{1\o N};1;k\right).}}
By repeated application of~\opid~we see that
\eqn\dprodp{D_N=D'_{\left({2N-5\o 2},{b_{N-1}\o N};1;k\right)}
D'_{\left({2N-7\o 2},{b_{N-2}\o N};1;k\right)}\ldots
D'_{\left(-{1\o 2},{b_1\o N};1;k\right)}}
where $(b_1,\ldots,b_{N-1})$ is any permutation of $(1,\ldots,N-1)$.
In particular, $b_1$ may take any value from 1 to $N-1$ from which it
immediately follows that
\eqn\fuchs{D_N e^{(N)}(k) = 0.}

\newsec{Discussion}

No finite order of perturbation theory can ever stand as the final
solution of any problem. However, such perturbation studies often
suggest many questions for further study and we conclude this note by
discussing some questions raised by this present study.

First it must be noted that the powers of the poles at any particular
location increase with the power of $k.$ This is an indication that
the location of the true singularity depends on $k$ and it would be
most desirable to find a perturbation theory directly for the location
of the singularities.

In terms of extending our results we remark that while
it may be cumbersome to extend the results for the non-integrable case
it is clear that for the integrable and superintegrable
cases a great deal more can
be computed for the perturbation expansion of the correlation
functions. It is to be hoped that further terms in these expansions
will produce conjectures such as~\conj~for the full correlation functions.

 Finally there is the question of whether or not the increased
analyticity which results from the the cancelation of poles  can be
taken as a sufficient condition for integrability which can be considered
as part of problem of the general relation between analyticity and
integrability.

 The construction of models which satisfy the
definition of integrability~\com~in terms of
commuting transfer matrices can be viewed as essentially a problem in
algebra. Indeed one of the major developments in algebra in the last
15 years is quantum groups which were invented specifically in order
to find classes of solutions to~\com. Analysis, on the other
hand seems to have only a marginal relation to~\com~and, in the
larger scheme of mathematics, algebra and analysis are almost
universally treated as separate disciplines.

A corresponding separation of disciplines has been traditional in
physics where statistical mechanics has relied heavily on algebra
while field theory has relied on analysis. Thus, for example, while
Onsager's solution of the Ising model~\rons~is totally
algebraic the renormalization theory of quantum electrodynamics is
almost exclusively concerned with subtractions of infinities in
integrals and seems to be totally concerned with analysis.

Thus 50 years ago integrability (algebra) and analyticity (analysis)
were seen to be totally separate. However, in the 50 years that have
followed these two totally separate computations there has been a
remarkable merging of field theory and statistical mechanics which can
be viewed as either the introduction of algebra into field theory or
the introduction of
analysis into statistical mechanics. Thus, as an example, at the
present day we have the isomorphism in two dimensions
of solvable statistical mechanical models (solved by algebra) with
conformal quantum field theory (where analysis appears in the very
statement of the subject).

If we take this merging of analyticity and integrability seriously,
however, then we are forced to conclude that the recent four
dimensional studies on $N=2$ Yang Mills theory~\rs-\rsw~which have
total reliance on analyticity properties should have an isomorphic counterpart
whose solution relies on algebra. The discovery of such an algebraic
counterpart (or the demonstration that it does not exist) to the work
of~\rs-\rsw~ will constitute an important advance in our
understanding of the relation of integrability and analyticity.
  \bigskip
  \noindent
  {\bf Acknowledgments.}
The authors wish to thank  Prof.~Gert Almkvist, Prof.~G.~von Gehlen,
Prof.~A.~Guttmann and Dr.~A.~Honecker for insightful discussions.
This work was partially supported by NSF grant DMR 9404747.

\appendix{A}{Proof of Trigonometric Identity~\trigidb}

To prove~\trigidb~we begin by using the symmetries of $f(N,j,k)$
and the cotangent function to write
\eqn\trigide{S(N) = 4 \sum_{j=1}^\ul \sum_{k=j+1}^{N-j-1}
P(N,j,k) \cot {\pi j \over N} \cot {\pi k \over N} +
4 \sum_{j=1}^\ul P(N,j,j) \cot^2 {\pi j \over N}.}
We recall that $P(N,j,k)$ was defined by~\pdef~in terms of a function
$Q(N,x)$ satisfying the symmetry property~\qsymm~and then
write $Q(N,x)$ so that the symmetry property automatically holds
\eqn\qdef{Q(N,x) = r(x)-r(1-x)}
where $r(x)$ is an arbitrary function whose $N$ dependence has been
suppressed for brevity.  Using this definition we expand~\trigide~as
\eqn\trigidf{\eqalign{S(N)
= & 8 \sum_{j=1}^\ul \sum_{k=j+1}^{N-j-1}
\left[ r\left({k+j \over N}\right)-r\left( {k \over N}\right)\right]
\cot {\pi j \over N} \cot {\pi k \over N} \cr
& + 4 \sum_{j=1}^\ul \left[ r\left( {2j \over N}\right)+r(0)+
2 r\left( {N-j \over N}\right)\right] \cot^2 {\pi j \over N}  \cr
&- 8 \sum_{j=1}^\ul \sum_{k=j+1}^{N-j-1}
\left[ r\left({N-k-j \over N}\right)-r\left( {N-k \over N}\right)\right]
\cot {\pi j \over N} \cot {\pi k \over N} \cr
& - 4 \sum_{j=1}^\ul \left[ r\left( {N-2j \over N}\right)+r(1)+
2 r\left( {j \over N}\right)\right] \cot^2 {\pi j \over N}}}
which may be rewritten as
\eqn\trigidg{S(N)= \sum_{a=0}^N C(N,a) r\left({a \over N}\right)=
\sum_{a=0}^\ul C(N,a) Q\left( N,{a \over N}\right)}
where
\eqn\czero{C(N,a)=\cases{4 \sum_{j=1}^\ul \cot^2 {\pi j \over N}& for $a=0$\cr
4C_1(N,a)+4C_2(N,a)& for $1 \le a \le \left\lfloor {N-1 \over 2}
\right\rfloor$\cr
-C(N,N-a)& for all $a$\cr}}
with
\eqn\cI{C_1(N,a)= 2 \sum_{j=1}^\ula \cot {\pi j \over N}
\cot {\pi (a-j) \over N}
-2 \sum_{j=1}^{a-1} \cot {\pi j \over N} \cot {\pi a \over N}
+\epsilon_a \cot^2 {\pi a \over 2N}}
and
\eqn\cII{\eqalign{C_2(N,a)=& -2 \sum_{j=1}^\ulNa \cot {\pi j \over N}
\cot {\pi (N-a-j) \over N}
+2 \sum_{j=1}^{a-1} \cot {\pi j \over N} \cot {\pi (N-a) \over N} \cr
&-\epsilon_{N-a} \cot^2 {\pi a \over 2N}-2\cot^2 {\pi a \o N}.}}
Here
\eqn\epsdef{\epsilon_a=\cases{1& if $a$ is even\cr
0& if $a$ is odd. \cr}}

The sum $C(N,0)$ may be rewritten as
\eqn\czeroa{C(N,0)= 2 \sum_{j=1}^{N-1} \cot^2 {\pi j \over N}}
which is recognized as the Dedekind sum~\rrade~and hence
\eqn\ded{C(N,0)={2(N-1)(N-2) \over 3}.}
To evaluate $C_1(N,a)$ we first write
\eqn\cIa{\eqalign{C_1(N,a)=&\sum_{j=1}^\ula \cot {\pi j \over N}
\cot {\pi (a-j) \over N}
+\sum_{j=\lla}^{a-1} \cot {\pi (a-j) \over N} \cot {\pi j \over N} \cr
&+\epsilon_a \cot^2 {\pi a \over 2N}
-2 \sum_{j=1}^{a-1} \cot {\pi j \over N} \cot {\pi a \over N} \cr
=&\sum_{j=1}^{a-1} \cot {\pi j \over N} \cot {\pi (a-j) \over N}
-2 \sum_{j=1}^{a-1} \cot {\pi j \over N} \cot {\pi a \over N} \cr
=&-\sum_{j=1}^{a-1} \cot {\pi j \over N} \cot {\pi (j+N-a) \over N}
-2 \cot {\pi (N-a) \over N} \sum_{j=1}^{N-a} \cot {\pi j \over N}.}}
Then let $\om=\exp\left({2\pi i \over N}\right)$ and write the first summand as
\eqn\cotcot{\eqalign{
&-\cot {\pi j \over N} \cot {\pi (j+N-a) \over N} = {\om^j+1 \o \om^j-1}\cdot
{\om^{j+N-a}+1 \o \om^{j+N-a}-1} \cr
&~~~=\left[1+{2 \o \om^j-1}\right]\left[1+{2 \o \om^{j+N-a}-1}\right] \cr
&~~~=\left[1+{2 \o \om^j-1}+{2 \o \om^{j+N-a}-1}+4{1 \o \om^j-1}\cdot
{1 \o \om^{j+N-a}-1} \right] \cr
&~~~=\biggl[1+{2 \o \om^j-1}+{2 \o \om^{j+N-a}-1}
+4\left({1 \o \om^j-1}-
{\om^{N-a} \o \om^{j+N-a}-1}\right){1 \o \om^{N-a}-1}\biggr] \cr
&~~~=\biggl[1+{2 \o (\om^j-1)(\om^{N-a}-1)}(\om^{N-a}-1+2)\cr
&~~~~~~~~~~~+{2 \o (\om^{j+N-a}-1)(\om^{N-a}-1)}
(\om^{N-a}-1-2\om^{N-a}) \biggr] \cr
&~~~=\left[ 1+2{\om^{N-a}+1 \o \om^{N-a}-1} \left({1 \o \om^j-1} -
{1 \o \om^{j+N-a}-1}\right)\right].}}
The sum partially telescopes
\eqn\cotsum{\eqalign{
&-\sum_{j=1}^{a-1} \cot {\pi j \o N} \cot {\pi (j+N-a) \o N} \cr
&~~~=(a-1) +2 {\om^{N-a}+1 \o \om^{N-a}-1}
\biggl[{1 \o \om^1 -1}+{1 \o \om^2 -1}+\cdots+{1 \o \om^{N-a} -1}\cr
&~~~~~~~~~~~~~~~~-{1 \o \om^{N-1} -1}-{1 \o \om^{N-2} -1}
-\cdots-{1 \o \om^a -1}
\biggr] \cr
&~~~=(a-1) +2 {\om^{N-a}+1 \o \om^{N-a}-1}
\left[{\om^1+1 \o \om^1 -1}+{\om^2+1 \o \om^2 -1}+\cdots+
{\om^{N-a}+1 \o \om^{N-a} -1} \right] \cr
&~~~=(a-1) +2 \cot {\pi (N-a) \o N} \sum_{j=1}^{N-a} \cot {\pi j \o N}.}}
It follows that
\eqn\cIb{C_1(N,a)=a-1.}

A parallel computation shows that
\eqn\cIIa{\eqalign{C_2(N,a) =&
\sum_{j=1}^{N-a-1} \cot {\pi j \over N} \cot {\pi (j+a) \over N}
-2 \cot {\pi a \over N} \sum_{j=1}^a \cot {\pi j \over N} \cr
=& -C_1(N,N-a)=-(N-a-1)}}
and therefore for $a\neq 0$
\eqn\cfinal{C(N,a)=-4(N-2a).}

\appendix{B}{Additional Identities Related to~\trigidb}

The cotangents on the left hand side of~\trigida~may be expanded in powers of
$\om=\exp\left({2\pi i \o N}\right)$ but the roots of unity all
cancel in the sum.
We expect that similar cancelations will occur when
the sum rule for the superintegrable correlation functions is
applied to orders higher than four.  Thus we have investigated
the generalization of~\trigida
\eqn\trigidh{S(N)=\sum_{j=1}^{N-1} \sum_{k=1}^{N-1} h_P(N,j,k)
\cot {\pi j \over N} \cot {\pi k \over N} =\hbox{polynomial in $N$}}
where
\eqn\hgen{h_P(N,j,k)=\cases{P(N,j,k)& if $1\leq j\leq k\leq N/2$\cr
h_P(N,k,j)& in general \cr
-h_P(N,N-j,k)& in general \cr}}
and where $P(N,j,k)$ is a polynomial in $N$, $j$ and $k$ of total degree
$d$.  In the case where
$N$ is prime, the only way for the roots of unity to vanish is for the
coefficients of $\om,\om^2,\ldots,\om^{N-1}$ all to be equal.  This
implies a system of equations for $P(N,j,k)$ which is overdetermined
when $N$ is sufficiently large.  Remarkably, there are many solutions
to this system and these solutions hold for all values of $N$ tested.
We conjecture that they are true for all $N$.

Up to the largest degree tested, $d=14$, the number of solutions
to the overdetermined system
is equal to the number of solutions to
\eqn\deg{d=3f+2n}
with $f$ a positive integer and $n$ a non-negative integer.  Thus
we make the further conjecture that there are infinitely many solutions
which can be organized into families, $f=1,2,3,\ldots$, whose
members, labeled by the index $n$, have degree given by~\deg.
Although there is immense freedom in combining solutions, it has
proven to be possible to write the solutions in such a form that
general formulas, given below, can be guessed in the cases $f=1,2$.
The solutions, $P_{f,n}(N,j,k)$, that we have found for $f=3,4$
are shown in Table~8~and the corresponding
sums, $S_{f,n}(N,j,k)$, are shown in Table~9.

The first family, $f=1$, comprises polynomials of the form~\pdef.
If in equation~\trigidb~we make the choice, $Q(N,x)=N^d q_d(x)$
with $q_d(x)$ a polynomial of degree $d$, then $S(N)$ is a polynomial in
$N$ of degree $d+2$.  The symmetry property on $Q(N,x)$,~\qsymm,~requires
$d$ to be odd and $d=1$ implies $P(N,j,k)=0$.  Thus these solutions have
the degree stated in~\deg.

The second family of solutions is given by
\eqn\trigidc{\eqalign{P_{2,n}(N,j,k) =& N^{2n+6} \sum_{h=0}^{n} \biggl[
\left( {j \over N} \right) ^{2n+3-2h} b_{nh} B_{2h+3} \left( {k \over N}
\right) + \cr
&\left( {j \over N} \right) ^{2n+2-2h} e_{nh} E_{2h+3} \left( {k \over N}
\right) \biggr]}}
where, letting $g(n) = 4^{n+1} - 1$,
\eqn\bedef{\eqalign{b_{nh} & = 2 {2n+3 \choose 2h} {1 \over (2h+1)(2h+2)(2h+3)}
{g(h) g(n-h) \over g(n)} \cr
e_{nh} & = 2 {2n+3 \choose 2h+1} {1 \over (2h+2)(2h+3)}
{g(h) 4^{n-h} \over g(n)},}}
and where $B_n(x)$ and $E_n(x)$ are the Bernoulli and Euler polynomials.
The sum in~\trigidc~may be performed using the identities~\ras
\eqn\beid{\eqalign{
B_n(x+h) & = \sum_{k=0}^{n} {n \choose k} B_k(x) h^{n-k} \cr
E_n(x+h) & = \sum_{k=0}^{n} {n \choose k} E_k(x) h^{n-k}}}
to obtain
\eqn\trigidd{\eqalign{
P_{2n} & = {N^{2n+6}\over(2n+6)(2n+5)(2n+4)(4^{n+1}-1)} \Biggl[
 (2 \cdot 4^{n+1}+1) B_{2n+6} \left( {k+j \over N} \right) - \cr
 &  (6 \cdot 4^{n+1}+1) B_{2n+6} \left( {k-j \over N} \right) +
  (4^{n+2}-1)   B_{2n+6} \left( {k \over N} \right) +
                B_{2n+6} \left( {k-2j \over N} \right) - \cr
 &  2 \cdot 4^{n+2}    \left( B_{2n+6} \left( {k+j/2 \over N} \right) -
                B_{2n+6} \left( {k-j/2 \over N} \right) \right) + \cr
 &  2 \cdot 4^{2n+4}   \left( B_{2n+6} \left( {k+j \over 2N} \right) +
                B_{2n+6} \left( {k-j \over 2N} \right) \right) -
  4^{n+3} (4^{n+2}-1) B_{2n+6} \left( {k \over 2N} \right) - \cr
 &  2 \cdot 4^{n+2}  \left( B_{2n+6} \left( {k+2j \over 2N} \right) +
                B_{2n+6} \left( {k-2j \over 2N} \right) \right) \Biggr] .}}
This expression differs from~\pdef~in that the Bernoulli polynomials
cannot be replaced by arbitrary polynomials satisfying $q(1-x)=-q(x)$.
A general form has not been found for the corresponding sums, $S_{2,n}(N)$.
The first five are shown in Table~7.

\vfill
\eject

{\bf Table~1}
The first nine orders of perturbation in $k$ for $N=3$ for the ground
state energy in the generic case. We use the variable $C_3=\cos(\phi/3).$

$$
{\ninepoint
\eqalign{e^{(3)}_0&=-{4 C_3\over {\sqrt 3}}~~~e{(3)}_1=0~~~
e^{(3)}_2={{-2}\over {3\,{\sqrt{3}}\,C_3}}\cr
\noalign{\smallskip}
e^{(3)}_3&={{-{\cos{\bar \phi}}}\over {9\,{\sqrt{3}}\,{{C_3}^2}}}\cr
\noalign{\smallskip}
e^{(3)}_4&={\sqrt{3}}\,\left( {1\over {162\,{{C_3}^3}}} +
    {4\over {81\,C_3}} -
    {4\over {81\,\left( -1 + 2\,C_3 \right) }} -
    {4\over {81\,\left( 1 + 2\,C_3 \right) }} \right)\cr
\noalign{\smallskip}
e^{(3)}_5&=-\left( {\sqrt{3}}\,\left( {{-{\cos{\bar \phi}}}\over
        {108\,{{C_3}^4}}} -
      {{4\,{\cos{\bar \phi}}}\over {81\,{{C_3}^2}}} +
      {{8\,{\cos{\bar \phi}}}\over {81\,\left( -1 + 2\,C_3 \right) }} -
      {{8\,{\cos{\bar \phi}}}\over {81\,\left( 1 + 2\,C_3 \right) }}
       \right)  \right) \cr
\noalign{\smallskip}
e^{(3)}_6
 &={\sqrt{3}}\,\biggl( {{-2}\over
      {729\,{{\left( -1 + 2\,C_3 \right) }^3}}}
  - {1\over {729\,{{\left( -1 + 2\,C_3 \right) }^2}}}
  + {{2\,\left( 1 - 54\,{{{\cos^2{\bar \phi}}}} \right) }\over
      {2187\,\left( -1 + 2\,C_3 \right) }} \cr
 &- {2\over {729\,{{\left( 1 + 2\,C_3 \right) }^3}}}
  + {1\over {729\,{{\left( 1 + 2\,C_3 \right) }^2}}}
  + {{2\,\left( 1 - 54\,{{{\cos^2{\bar \phi}}}} \right) }\over
      {2187\,\left( 1 + 2\,C_3 \right) }} \cr
 &+ {{11 + 24\,{{{\cos^2{\bar \phi}}}}}\over {11664\,{{C_3}^5}}}
  + {{5 + 54\,{{{\cos^2{\bar \phi}}}}}\over
      {4374\,{{C_3}^3}}}
  + {{2\,\left( -1 + 54\,{{{\cos^2{\bar \phi}}}} \right) }\over
      {2187\,C_3}}
 \biggr) \cr
\noalign{\smallskip}
e^{(3)}_7
 &=-{\sqrt{3}}\,\biggl( {{{\cos{\bar \phi}}}\over {972\,{{C_3}^6}}} +
      {{263\,{\cos{\bar \phi}}}\over {26244\,{{C_3}^4}}} +
      {{1022\,{\cos{\bar \phi}}}\over {19683\,{{C_3}^2}}} \cr
   &+ {{{\cos{\bar \phi}}}\over
        {2187\,{{\left( -1 + 2\,C_3 \right) }^4}}}
  + {{64\,{\cos{\bar \phi}}}\over
        {6561\,{{\left( -1 + 2\,C_3 \right) }^3}}} -
      {{119\,{\cos{\bar \phi}}}\over
        {39366\,{{\left( -1 + 2\,C_3 \right) }^2}}} -
      {{49\,{\cos{\bar \phi}}}\over
        {486\,\left( -1 + 2\,C_3 \right) }} \cr
   &+ {{{\cos{\bar \phi}}}\over
        {2187\,{{\left( 1 + 2\,C_3 \right) }^4}}}
  - {{64\,{\cos{\bar \phi}}}\over
        {6561\,{{\left( 1 + 2\,C_3 \right) }^3}}} -
      {{119\,{\cos{\bar \phi}}}\over
        {39366\,{{\left( 1 + 2\,C_3 \right) }^2}}} +
      {{49\,{\cos{\bar \phi}}}\over {486\,\left( 1 + 2\,C_3 \right) }}
       \biggr)  \cr
\noalign{\smallskip}
e^{(3)}_8
 &={\sqrt{3}}\,\biggl( {{-1}\over
      {2187\,{{\left( -1 + 2\,C_3 \right) }^5}}}
  + {{-1 - 2\,{{{\cos^2{\bar \phi}}}}}\over
      {2187\,{{\left( -1 + 2\,C_3 \right) }^4}}}
  + {{55 - 888\,{{{\cos^2{\bar \phi}}}}}\over
      {78732\,{{\left( -1 + 2\,C_3 \right) }^3}}} \cr
 &+ {{-451 + 1384\,{{{\cos^2{\bar \phi}}}}}\over
      {157464\,{{\left( -1 + 2\,C_3 \right) }^2}}}
  + {{2759 + 15264\,{{{\cos^2{\bar \phi}}}}}\over
      {78732\,\left( -1 + 2\,C_3 \right) }}
  - {1\over {2187\,{{\left( 1 + 2\,C_3 \right) }^5}}}
  + {{1 + 2\,{{{\cos^2{\bar \phi}}}}}\over
      {2187\,{{\left( 1 + 2\,C_3 \right) }^4}}} \cr
 &+ {{55 - 888\,{{{\cos^2{\bar \phi}}}}}\over
      {78732\,{{\left( 1 + 2\,C_3 \right) }^3}}}
  + {{451 - 1384\,{{{\cos^2{\bar \phi}}}}}\over
      {157464\,{{\left( 1 + 2\,C_3 \right) }^2}}}
  + {{2759 + 15264\,{{{\cos^2{\bar \phi}}}}}\over
      {78732\,\left( 1 + 2\,C_3 \right) }}
  - {{1715\,C_3}\over {2187\,\left( -3 + 28\,{{C_3}^2} \right) }} \cr
 &+ {{-381 - 1936\,{{{\cos^2{\bar \phi}}}}}\over {1679616\,{{C_3}^7}}}
  + {{-801 - 6208\,{{{\cos^2{\bar \phi}}}}}\over {629856\,{{C_3}^5}}}
  + {{-709 - 7880\,{{{\cos^2{\bar \phi}}}}}\over {157464\,{{C_3}^3}}}
  + {{-277 - 7632\,{{{\cos^2{\bar \phi}}}}}\over {39366\,C_3}}
 \biggr) \cr}}$$
\vfill
\eject
$${\ninepoint
\eqalign{
e^{(3)}_9
 &=- {\sqrt{3}}\,\biggl( {{5\,{\cos{\bar \phi}}}\over
        {26244\,{{\left( -1 + 2\,C_3 \right) }^6}}}
    + {{17\,{\cos{\bar \phi}}}\over
        {8748\,{{\left( -1 + 2\,C_3 \right) }^5}}}
    + {{61\,{\cos{\bar \phi}} + 432\,{{{\cos^3{\bar \phi}}}}}\over
        {944784\,{{\left( -1 + 2\,C_3 \right) }^4}}} \cr
   &+ {{-11257\,{\cos{\bar \phi}} + 6048\,{{{\cos^3{\bar \phi}}}}}\over
        {1417176\,{{\left( -1 + 2\,C_3 \right) }^3}}}
    + {{105767\,{\cos{\bar \phi}} - 24912\,{{{\cos^3{\bar \phi}}}}}\over
        {5668704\,{{\left( -1 + 2\,C_3 \right) }^2}}}
    + {{-34607\,{\cos{\bar \phi}} - 59184\,{{{\cos^3{\bar \phi}}}}}\over
        {629856\,\left( -1 + 2\,C_3 \right) }} \cr
   &+ {{5\,{\cos{\bar \phi}}}\over
        {26244\,{{\left( 1 + 2\,C_3 \right) }^6}}}
    - {{17\,{\cos{\bar \phi}}}\over
        {8748\,{{\left( 1 + 2\,C_3 \right) }^5}}}
    + {{61\,{\cos{\bar \phi}} + 432\,{{{\cos^3{\bar \phi}}}}}\over
        {944784\,{{\left( 1 + 2\,C_3 \right) }^4}}}
    + {{11257\,{\cos{\bar \phi}} - 6048\,{{{\cos^3{\bar \phi}}}}}\over
        {1417176\,{{\left( 1 + 2\,C_3 \right) }^3}}} \cr
   &+ {{105767\,{\cos{\bar \phi}} - 24912\,{{{\cos^3{\bar \phi}}}}}\over
        {5668704\,{{\left( 1 + 2\,C_3 \right) }^2}}}
    + {{34607\,{\cos{\bar \phi}} + 59184\,{{{\cos^3{\bar \phi}}}}}\over
        {629856\,\left( 1 + 2\,C_3 \right) }}
    + {{24010\,{\cos{\bar \phi}}}\over
        {19683\,\left( -3 + 28\,{{C_3}^2} \right) }} \cr
   &+ {{133\,{\cos{\bar \phi}} + 192\,{{{\cos^3{\bar \phi}}}}}\over
        {839808\,{{C_3}^8}}}
    + {{13\,\left( 11\,{\cos{\bar \phi}} + 52\,{{{\cos^3{\bar \phi}}}} \right)
          }\over {314928\,{{C_3}^6}}}
    + {{-575\,{\cos{\bar \phi}} + 3696\,{{{\cos^3{\bar \phi}}}}}\over
        {314928\,{{C_3}^4}}} \cr
   &+ {{-9007\,{\cos{\bar \phi}} + 17424\,{{{\cos^3{\bar \phi}}}}}\over
        {354294\,{{C_3}^2}}}
 \biggr)  \cr}}$$
\vfill
\eject
{\bf Table~2}
The first eight orders of perturbation in  $k$ for $N=4$ for the
ground state energy in the generic case. We use the variable
$C_4={\sqrt 2}\cos(\phi /2).$
$$\ninepoint
\eqalign{e^{(4)}_0&=-1-2C_4,~~~e^{(4)}_1=0~~~
e^{(4)}_2={{-1}\over {8\,C_4}} - {1\over {1 + C_4}}\cr
\noalign{\smallskip}
e^{(4)}_3&=-{{\cos{\bar \phi}}\over {4\,C_4}} -
  {{{\cos{\bar \phi}}}\over {4\,{{\left( 1 + C_4 \right) }^2}}} +
  {{{\cos{\bar \phi}}}\over {4\,\left( 1 + C_4 \right)}}\cr
\noalign{\smallskip}
e^{(4)}_4&={{-1}\over {512\,{{C_4}^3}}} + {1\over {16\,{{C_4}^2}}} +
  {{5 - 2\,{{{\cos^2{\bar \phi}}}}}\over {16\,C_4}} +
  {1\over {12\,{{\left( 1 + C_4 \right) }^3}}} +
  {{8 + 9\,{{{\cos^2{\bar \phi}}}}}\over
    {72\,{{\left( 1 + C_4 \right) }^2}}}\cr
& +  {{20 + 27\,{{{\cos^2{\bar \phi}}}}}\over
    {216\,\left( 1 + C_4 \right) }} -
  {{875}\over {432\,\left( -1 + 5\,C_4 \right) }}\cr
\noalign{\smallskip}
e^{(4)}_5
 &={{3\,{\cos{\bar \phi}}}\over {128\,{{C_4}^3}}} +
  {{19\,{\cos{\bar \phi}}}\over {64\,{{C_4}^2}}} +
  {{149\,{\cos{\bar \phi}}}\over {128\,C_4}} +
  {{61\,{\cos{\bar \phi}}}\over {576\,{{\left( 1 + C_4 \right) }^4}}} +
  {{55\,{\cos{\bar \phi}}}\over {864\,{{\left( 1 + C_4 \right) }^3}}} \cr
 &+ {{{\cos{\bar \phi}}}\over
  {1152\,{{\left( 1 + C_4 \right) }^2}}} -
  {{277\,{\cos{\bar \phi}}}\over {31104\,\left( 1 + C_4 \right) }} -
  {{1225\,{\cos{\bar \phi}}}\over
    {648\,{{\left( -1 + 5\,C_4 \right) }^2}}} -
  {{89825\,{\cos{\bar \phi}}}\over {15552\,\left( -1 + 5\,C_4 \right) }}
\cr
\noalign{\smallskip}
e^{(4)}_6
 &={{2025}\over {8192\,\left( -1 - 3\,C_4 \right) }} -
  {1\over {8192\,{{C_4}^5}}} + {3\over {512\,{{C_4}^4}}} +
  {{-7 + 52\,{{{\cos^2{\bar \phi}}}}}\over {1024\,{{C_4}^3}}} +
  {{3\,\left( -23 + 48\,{{{\cos^2{\bar \phi}}}} \right) }\over
    {512\,{{C_4}^2}}} \cr
 &+ {{-303 + 812\,{{{\cos^2{\bar \phi}}}}}\over
    {1024\,C_4}} + {{155 - 44\,{{{\cos^2{\bar \phi}}}}}\over
    {3456\,{{\left( 1 + C_4 \right) }^5}}} +
  {{-33 - 394\,{{{\cos^2{\bar \phi}}}}}\over
    {3456\,{{\left( 1 + C_4 \right) }^4}}} +
  {{-253 - 3911\,{{{\cos^2{\bar \phi}}}}}\over
    {20736\,{{\left( 1 + C_4 \right) }^3}}} \cr
 &+ {{-15832 - 72167\,{{{\cos^2{\bar \phi}}}}}\over
    {373248\,{{\left( 1 + C_4 \right) }^2}}} +
  {{75203 - 236166\,{{{\cos^2{\bar \phi}}}}}\over
    {1492992\,\left( 1 + C_4 \right) }} -
  {{8575\,\left( 5 + 4\,{{{\cos^2{\bar \phi}}}} \right) }\over
    {15552\,{{\left( -1 + 5\,C_4 \right) }^3}}} \cr
 &+ {{175\,\left( 8345 - 12704\,{{{\cos^2{\bar \phi}}}} \right) }\over
    {746496\,{{\left( -1 + 5\,C_4 \right) }^2}}} +
  {{25\,\left( 727595 - 758184\,{{{\cos^2{\bar \phi}}}} \right) }\over
    {5971968\,\left( -1 + 5\,C_4 \right) }} +
  {{9\,\left( 1 + C_4 \right) }\over
    {16\,\left( 1 - 2\,C_4 - 2\,{{C_4}^2} \right) }} \cr
\noalign{\smallskip}
e^{(4)}_7
 &={{5\,{\cos{\bar \phi}}}\over {4096\,{{C_4}^5}}} -
  {{31\,{\cos{\bar \phi}}}\over {1024\,{{C_4}^4}}} -
  {{1301\,{\cos{\bar \phi}} - 96\,{{{\cos^3{\bar \phi}}}}}\over
    {4096\,{{C_4}^3}}} - {{309\,{\cos{\bar \phi}} +
      20\,{{{\cos^3{\bar \phi}}}}}\over {256\,{{C_4}^2}}} \cr
 -&{{3855\,{\cos{\bar \phi}} + 2944\,{{{\cos^3{\bar \phi}}}}}\over
    {4096\,C_4}} - {{101\,{\cos{\bar \phi}} +
      960\,{{{\cos^3{\bar \phi}}}}}\over
    {41472\,{{\left( 1 + C_4 \right) }^6}}} -
  {{13385\,{\cos{\bar \phi}} + 5904\,{{{\cos^3{\bar \phi}}}}}\over
    {124416\,{{\left( 1 + C_4 \right) }^5}}} \cr
 -& {{124655\,{\cos{\bar \phi}} + 83952\,{{{\cos^3{\bar \phi}}}}}\over
    {1492992\,{{\left( 1 + C_4 \right) }^4}}} -
  {{96257\,{\cos{\bar \phi}} + 14169\,{{{\cos^3{\bar \phi}}}}}\over
    {559872\,{{\left( 1 + C_4 \right) }^3}}} -
  {{-4392121\,{\cos{\bar \phi}} - 488304\,{{{\cos^3{\bar \phi}}}}}\over
    {26873856\,{{\left( 1 + C_4 \right) }^2}}} \cr
 -& {{11116697\,{\cos{\bar \phi}} -
     1116063\,{{{\cos^3{\bar \phi}}}}}\over
    {20155392\,\left( 1 + C_4 \right) }} +
  {{1539\,{\cos{\bar \phi}}}\over
    {16384\,{{\left( 1 + 3\,C_4 \right) }^2}}} -
  {{25839\,{\cos{\bar \phi}}}\over
    {32768\,\left( 1 + 3\,C_4 \right) }} \cr
 -& {{1860775\,{\cos{\bar \phi}}}\over
    {186624\,{{\left( -1 + 5\,C_4 \right) }^4}}} -
  {{1225\,\left( 4103\,{\cos{\bar \phi}} + 2800\,{{{\cos^3{\bar \phi}}}}
         \right) }\over {746496\,{{\left( -1 + 5\,C_4 \right)}^3}}}\cr
 -& {{25\,\left( -47222099\,{\cos{\bar \phi}} +
        621600\,{{{\cos^3{\bar \phi}}}} \right) }\over
    {35831808\,{{\left( -1 + 5\,C_4 \right) }^2}}}
 + {{25\,\left( 432042539\,{\cos{\bar \phi}} +
        85572000\,{{{\cos^3{\bar \phi}}}} \right) }\over
    {644972544\,\left( -1 + 5\,C_4 \right) }}\cr
 -&  {{51\,\left( 2\,{\cos{\bar \phi}} + {\cos{\bar \phi}}\,C_4 \right)}
     \over {16\,\left( -1 + 2\,C_4 + 2\,{{C_4}^2} \right) }} \cr} $$
\vfill
\eject
$$\ninepoint
\eqalign{e^{(4)}_8
 &={{81\,\left( 199 + 32\,{{{\cos^2{\bar \phi}}}} \right) }\over
    {131072\,{{\left( -1 - 3\,C_4 \right) }^3}}} +
  {{81\,\left( -803 + 7760\,{{{\cos^2{\bar \phi}}}} \right) }\over
    {1048576\,{{\left( -1 - 3\,C_4 \right) }^2}}} +
  {{81\,\left( -20863 + 102076\,{{{\cos^2{\bar \phi}}}} \right) }\over
    {13631488\,\left( -1 - 3\,C_4 \right) }}\cr
& - {{25}\over {2097152\,{{C_4}^7}}}
 + {9\over {16384\,{{C_4}^6}}} +
  {{-151 - 158\,{{{\cos^2{\bar \phi}}}}}\over {16384\,{{C_4}^5}}} +
  {{-519 - 3008\,{{{\cos^2{\bar \phi}}}}}\over {16384\,{{C_4}^4}}}\cr
& +
  {{359 - 15946\,{{{\cos^2{\bar \phi}}}} - 32\,{{{\cos^4{\bar \phi}}}}}\over
    {16384\,{{C_4}^3}}} \cr
 &+ {{7923 - 43936\,{{{\cos^2{\bar \phi}}}} -
      2048\,{{{\cos^4{\bar \phi}}}}}\over {16384\,{{C_4}^2}}} +
  {{76123 + 14074\,{{{\cos^2{\bar \phi}}}} -
      10816\,{{{\cos^4{\bar \phi}}}}}\over {16384\,C_4}} \cr
 &+ {{-53297 + 31488\,{{{\cos^2{\bar \phi}}}} -
      4608\,{{{\cos^4{\bar \phi}}}}}\over
    {1990656\,{{\left( 1 + C_4 \right) }^7}}} +
  {{-555061 - 193632\,{{{\cos^2{\bar \phi}}}} +
      138240\,{{{\cos^4{\bar \phi}}}}}\over
    {23887872\,{{\left( 1 + C_4 \right) }^6}}} \cr
 &+ {{-1933609 + 12161664\,{{{\cos^2{\bar \phi}}}} +
      2257920\,{{{\cos^4{\bar \phi}}}}}\over
    {95551488\,{{\left( 1 + C_4 \right) }^5}}}\cr
& +{{85101083 - 62007936\,{{{\cos^2{\bar \phi}}}} +
      179776512\,{{{\cos^4{\bar \phi}}}}}\over
    {3439853568\,{{\left( 1 + C_4 \right) }^4}}} \cr
 &+ {{-2291995019 + 25564353792\,{{{\cos^2{\bar \phi}}}} +
      3332192256\,{{{\cos^4{\bar \phi}}}}}\over
    {41278242816\,{{\left( 1 + C_4 \right) }^3}}} \cr
 &+ {{9085861571 - 47135217152\,{{{\cos^2{\bar \phi}}}} +
      5619560448\,{{{\cos^4{\bar \phi}}}}}\over
    {55037657088\,{{\left( 1 + C_4 \right) }^2}}} \cr
 &+ {{-917428110137 + 7122467770368\,{{{\cos^2{\bar \phi}}}} +
      222666719232\,{{{\cos^4{\bar \phi}}}}}\over
    {1981355655168\,\left( 1 + C_4 \right) }} -
  {{420175\,\left( 2 + 3\,{{{\cos^2{\bar \phi}}}} \right) }\over
    {69984\,{{\left( -1 + 5\,C_4 \right) }^5}}} \cr
 &+{{8575\,\left( -3751 + 21936\,{{{\cos^2{\bar \phi}}}} \right) }\over
    {26873856\,{{\left( -1 + 5\,C_4 \right) }^4}}} +
  {{175\,\left( 161445421 + 287088144\,{{{\cos^2{\bar \phi}}}} -
        35280000\,{{{\cos^4{\bar \phi}}}} \right) }\over
    {2579890176\,{{\left( -1 + 5\,C_4 \right) }^3}}} \cr
 &+ {{175\,\left( -835186517 + 7197054312\,{{{\cos^2{\bar \phi}}}} +
        37920000\,{{{\cos^4{\bar \phi}}}} \right) }\over
    {10319560704\,{{\left( -1 + 5\,C_4 \right) }^2}}} \cr
 &+
  {{125\,\left( 1145946953233 + 5452562686272\,{{{\cos^2{\bar \phi}}}} +
        282187776000\,{{{\cos^4{\bar \phi}}}} \right) }\over
    {12878811758592\,\left( -1 + 5\,C_4 \right) }} \cr
 &- {{29176875}\over {33554432\,\left( 1 + 5\,C_4 \right) }} +
  {{9\,\left( 4 + 3\,C_4 \right) }\over
    {128\,{{\left( 1 - 2\,C_4 - 2\,{{C_4}^2} \right) }^3}}} -
  {{3\,\left( 81 + 46\,C_4 \right) }\over
    {512\,{{\left( 1 - 2\,C_4 - 2\,{{C_4}^2} \right) }^2}}} \cr
 &+ {{-19777 + 197496\,{{{\cos^2{\bar \phi}}}} - 22351\,C_4 +
      197496\,{{{\cos^2{\bar \phi}}}}\,C_4}\over
    {6656\,\left( 1 - 2\,C_4 - 2\,{{C_4}^2} \right) }} -
  {{27\,\left( 48104891 + 84022143\,C_4 \right) }\over
    {16777216\,\left( -1 + 8\,C_4 + 17\,{{C_4}^2} \right) }} \cr}$$
\vfill
\eject
{\bf Table~3}
The first 5 orders of perturbation in $k$ for $N=5$ for the ground
state energy in the generic case. For the fourth and fifth orders only
the terms proportional to $\cos^2{\bar \phi}$ and $\cos^3{\bar \phi}$
respectively are given. We use the variable
$C_5={\cos(\phi/5)\over\sin\pi/5}.$

$$\ninepoint
\eqalign{e^{(5)}_0&=-(5-{\sqrt 5})C_5^3+(7-{\sqrt 5})C_5,~~~e^{(5)}_1=0\cr
e^{(5)}_2&=\left( 5 + {\sqrt{5}} \right) \,
  \left( {{-\left( 3 + {\sqrt{5}} \right) }\over
      {50\,\left( -1 + C_5 \right) }} +
    {{9 + {\sqrt{5}}}\over {50\,C_5}} -
    {{3 + {\sqrt{5}}}\over {50\,\left( 1 + C_5 \right) }} +
    {{\left( -3 + {\sqrt{5}} \right) \,C_5}\over
      {50\,\left( -2 + {{C_5}^2} \right) }} \right)\cr
\noalign{\smallskip}
e^{(5)}_3&=
 {{\sqrt{5 + {\sqrt{5}}}}\over {-5 + 3\,{\sqrt{5}}}}
 \,\biggl( {{-\left( \left( 5 + 3\,{\sqrt{5}} \right)\,
             {\cos{\bar \phi}} \right) }\over
         {125\,{\sqrt{2}}\,{{\left( -1 + C_5 \right) }^2}}} +
       {{\left( 35 + 13\,{\sqrt{5}} \right) \,{\cos{\bar \phi}}}\over
         {125\,{\sqrt{2}}\,\left( -1 + C_5 \right) }} -
       {{{\sqrt{2}}\,\left( 25 + 6\,{\sqrt{5}} \right) \,
           {\cos{\bar \phi}}}\over {125\,{{C_5}^2}}}\cr
 -&{{\left( 5 + 3\,{\sqrt{5}} \right) \,{\cos{\bar \phi}}}\over
         {125\,{\sqrt{2}}\,{{\left( 1 + C_5 \right) }^2}}}
 - {{\left( 35 + 13\,{\sqrt{5}} \right) \,{\cos{\bar \phi}}}\over
         {125\,{\sqrt{2}}\,\left( 1 + C_5 \right) }} +
       {{2\,{\sqrt{2}}\,\left( -5 + 2\,{\sqrt{5}} \right) \,
           {\cos{\bar \phi}}}\over
         {125\,{{\left( -2 + {{C_5}^2} \right) }^2}}} -
       {{{\sqrt{2}}\,\left( 5 + 4\,{\sqrt{5}} \right) \,
           {\cos{\bar \phi}}}\over {125\,\left( -2 + {{C_5}^2} \right) }}
        \biggr)  \cr
\noalign{\smallskip}
e_4^{(5)}
 &={{{\cos^2{\bar \phi}}}}\,\biggl( {{110 + 50\,{\sqrt{5}}}\over
      {3125\,{{\left( -1 + C_5 \right) }^2}}} -
    {{4\,\left( 15 + 7\,{\sqrt{5}} \right) }\over
      {625\,\left( -1 + C_5 \right) }} +
    {{260 + 120\,{\sqrt{5}}}\over {3125\,{{C_5}^3}}} +
    {{630 + 290\,{\sqrt{5}}}\over {3125\,C_5}} \cr
 &-{{110 + 50\,{\sqrt{5}}}\over {3125\,{{\left( 1 + C_5 \right) }^2}}} -
    {{4\,\left( 15 + 7\,{\sqrt{5}} \right) }\over
      {625\,\left( 1 + C_5 \right) }} +
    {{\left( -40 - 20\,{\sqrt{5}} \right) \,C_5}\over
      {3125\,{{\left( -2 + {{C_5}^2} \right) }^2}}} -
    {{\left( 30 + 10\,{\sqrt{5}} \right) \,C_5}\over
      {3125\,\left( -2 + {{C_5}^2} \right) }} \biggr)+\cdots \cr
\noalign{\smallskip}
e_5^{(5)}
 &=\left( {{-52\,{\sqrt{10 - 2\,{\sqrt{5}}}}}\over {3125}} -
    {{116\,{\sqrt{10 - 2\,{\sqrt{5}}}}}\over {3125\,{\sqrt{5}}}} \right) \,
  {{{\cos^3{\bar \phi}}}}\,\biggl( {1\over
      {4\,{{\left( -1 + C_5 \right) }^2}}} -
    {1\over {4\,\left( -1 + C_5 \right) }} \cr
 &+ {1\over {4\,{{C_5}^4}}} +
    {3\over {4\,{{C_5}^2}}} + {1\over {4\,{{\left( 1 + C_5 \right) }^2}}} +
    {1\over {4\,\left( 1 + C_5 \right) }} +
    {1\over {4\,{{\left( -2 + {{C_5}^2} \right) }^2}}} -
    {3\over {4\,\left( -2 + {{C_5}^2} \right) }} \biggr) \cr} $$

\vfill
\eject
{\bf Table~4}
The first 6 orders of perturbation in $k$ for $N=6$ for the ground state
energy in the generic case. For the sixth order only the term
proportional to $\cos^4{\bar \phi}$ is given. We use the variable $C_6={\sqrt
3}\cos(\phi /3).$

$$\ninepoint
\eqalign{e^{(6)}_0&=3-{4\over 3}C_6-{8\over 3}C_6^2,~~~e^{(6)}_1=0\cr
\noalign{\smallskip}
e_2^{(6)}&={{-2}\over {15\,\left( -1 + C_6 \right) }} - {2\over {C_6}} +
  {{64}\over {15\,\left( 3 + 2\,C_6 \right) }} -
  {3\over 4 (-9 + 8\,{{C_6}^2})} \cr
\noalign{\smallskip}
e_3^{(6)}
 &={{-{\cos{\bar \phi}}}\over
    {75\,{\sqrt{3}}\,{{\left( -1 + C_6 \right) }^2}}} +
  {{94\,{\cos{\bar \phi}}}\over
    {375\,{\sqrt{3}}\,\left( -1 + C_6 \right) }} -
  {{{\cos{\bar \phi}}}\over {{\sqrt{3}}\,{{C_6}^2}}} +
  {{8\,{\cos{\bar \phi}}}\over {3\,{\sqrt{3}}\,C_6}} \cr
 &- {{484\,{\cos{\bar \phi}}}\over
    {75\,{\sqrt{3}}\,{{\left( 3 + 2\,C_6 \right) }^2}}} -
  {{132\,{\sqrt{3}}\,{\cos{\bar \phi}}}\over
    {125\,\left( 3 + 2\,C_6 \right) }} -
  {{32\,{\cos{\bar \phi}}\,C_6 }\over
    {3\,{\sqrt{3}}\,\left( -9 + 8\,{{C_6}^2} \right) }} \cr
\noalign{\smallskip}
e_4^{(6)}
 &={1\over {2250\,{{\left( -1 + C_6 \right) }^3}}} +
  {{667 + 360\,{{{\cos^2{\bar \phi}}}}}\over
    {15000\,{{\left( -1 + C_6 \right) }^2}}} +
  {{216617 + 527040\,{{{\cos^2{\bar \phi}}}}}\over
    {450000\,\left( -1 + C_6 \right) }} \cr
 &+ {{11}\over {30\,{{C_6}^3}}} +
  {{2\,\left( -39 + 50\,{{{\cos^2{\bar \phi}}}} \right) }\over
    {225\,{{C_6}^2}}} - {{-3691 + 5000\,{{{\cos^2{\bar \phi}}}}}\over
    {6750\,C_6}} \cr
 &+ {{2\,\left( -2381 + 3060\,{{{\cos^2{\bar \phi}}}} \right)}
     \over {3375\,{{\left( 3 + 2\,C_6 \right) }^3}}} +
  {{-1620919 + 579960\,{{{\cos^2{\bar \phi}}}}}\over
    {911250\,{{\left( 3 + 2\,C_6 \right) }^2}}} -
  {{708766379 + 423636480\,{{{\cos^2{\bar \phi}}}}}\over
    {492075000\,\left( 3 + 2\,C_6 \right) }} \cr
 &-
  {{1294139}\over {243000\,\left( -15 + 14\,C_6 \right) }} -
  {{2\,\left( 3 + C_6 \right) }\over
    {99\,\left( -4 + C_6 + 2\,{{C_6}^2} \right) }} -
  {{4\,\left( 9 + 8\,C_6 \right) }\over
    {297\,\left( -3 + 4\,{{C_6}^2} \right) }} \cr
 &-
  {{27}\over {64\,{{\left( -9 + 8\,{{C_6}^2} \right) }^3}}} +
  {{4\,\left( -9 + 16\,C_6 \right) }\over
    {9\,{{\left( -9 + 8\,{{C_6}^2} \right) }^2}}} -
  {{44\,\left( -11 + 36\,{{{\cos^2{\bar \phi}}}} + 8\,C_6 \right) }\over
    {81\,\left( -9 + 8\,{{C_6}^2} \right) }} \cr
 &+
  {{9878103 - 19929087\,C_6 + 7810406\,{{C_6}^2}}\over
    {314928\,\left( 27 - 36\,C_6 - 33\,{{C_6}^2} + 38\,{{C_6}^3} \right) }}
 \cr}$$
\vfill
\eject
$$\ninepoint
\eqalign{
e_5^{(6)}
 &={{{\cos{\bar \phi}}}\over
    {1000\,{\sqrt{3}}\,{{\left( -1 + C_6 \right) }^3}}} +
  {{193\,{\cos{\bar \phi}}}\over
    {10000\,{\sqrt{3}}\,{{\left( -1 + C_6 \right) }^2}}} +
  {{{\cos{\bar \phi}}\,\left( -242459 + 194400\,{{{\cos^2{\bar \phi}}}}
         \right) }\over {600000\,{\sqrt{3}}\,\left( -1 + C_6 \right) }} \cr
 &+
  {{5427\,{\sqrt{3}}\,{\cos{\bar \phi}}}\over {160\,{{C_6}^3}}} -
  {{126279\,{\sqrt{3}}\,{\cos{\bar \phi}}}\over {800\,{{C_6}^2}}} -
  {{3\,{\sqrt{3}}\,{\cos{\bar \phi}}\,
      \left( -185083 + 11250\,{{{\cos^2{\bar \phi}}}} \right) }\over
    {2000\,C_6}} \cr
 &- {{81\,{\sqrt{3}}\,{\cos{\bar \phi}}\,
      \left( -65789 + 5400\,{{{\cos^2{\bar \phi}}}} \right) }\over
    {1000\,{{\left( 3 + 2\,C_6 \right) }^3}}} +
  {{9\,{\sqrt{3}}\,{\cos{\bar \phi}}\,
      \left( -842051 + 361575\,{{{\cos^2{\bar \phi}}}} \right) }\over
    {2500\,{{\left( 3 + 2\,C_6 \right) }^2}}} \cr
 &+
  {{{\cos{\bar \phi}}\,\left( -2735083651 +
        191181600\,{{{\cos^2{\bar \phi}}}} \right) }\over
    {800000\,{\sqrt{3}}\,\left( 3 + 2\,C_6 \right) }} +
  {{8349\,{\sqrt{3}}\,{\cos{\bar \phi}}}\over
    {32000\,\left( -15 + 14\,C_6 \right) }}\cr
& - {{469\,{\cos{\bar \phi}}\,\left( -302 + 255\,C_6 \right) }\over
    {186\,{\sqrt{3}}\,\left( -4 + C_6 + 2\,{{C_6}^2} \right) }}
+{{{\sqrt{3}}\,{\cos{\bar \phi}}\,\left( -229 + 264\,C_6 \right) }\over
    {8\,\left( -3 + 4\,{{C_6}^2} \right) }}\cr
& +{{3645\,{\sqrt{3}}\,{\cos{\bar \phi}}\,\left( -123 + 116\,C_6 \right) }\over
      {1024\,{{\left( -9 + 8\,{{C_6}^2} \right) }^3}}} -
  {{27\,{\sqrt{3}}\,{\cos{\bar \phi}}\,
      \left( -83589 + 79246\,C_6 \right) }\over
    {512\,{{\left( -9 + 8\,{{C_6}^2} \right) }^2}}} \cr
 &-
  {{9\,{\sqrt{3}}\,{\cos{\bar \phi}}\,
      \left( 8025965 - 2928384\,{{{\cos^2{\bar \phi}}}} - 8380680\,C_6 +
        2761728\,{{{\cos^2{\bar \phi}}}}\,C_6 \right) }\over
    {31744\,\left( -9 + 8\,{{C_6}^2} \right) }} \cr
 &+
  {{{\cos{\bar \phi}}\,\left( 22455 - 53739\,C_6 + 29648\,{{C_6}^2} \right)}
     \over {4\,{\sqrt{3}}\,\left( 27 - 36\,C_6 - 33\,{{C_6}^2} +
        38\,{{C_6}^3} \right) }} \cr
\noalign{\smallskip}
e_6^{(6)}
 &={{{\cos^4{\bar \phi}}}}\,\biggl( {{27}\over
      {625\,{{\left( -1 + C_6 \right) }^2}}} +
    {{1674}\over {3125\,\left( -1 + C_6 \right) }} +
    {1\over {27\,{{C_6}^2}}} - {2\over {81\,C_6}} \cr
 &- {{16}\over {75\,{{\left( 3 + 2\,C_6 \right) }^4}}} -
    {{64}\over {125\,{{\left( 3 + 2\,C_6 \right) }^3}}} -
    {{13312}\over {16875\,{{\left( 3 + 2\,C_6 \right) }^2}}}\cr
 &- {{258688}\over {253125\,\left( 3 + 2\,C_6 \right) }} -
    {{256}\over {27\,\left( -9 + 8\,{{C_6}^2} \right) }} \biggr)+\cdots \cr} $$

\vfill
\eject

{\bf Table~5}

The results of specializing the generic results of the perturbation
theory for the ground state energy for $N=3,4,5$ and $6$ to
the integrable manifold \int~where $\cos\phi=k\cos{\bar\phi}.$

$$\ninepoint
\eqalign{e^{(3)}=&{{-4C_3}\over {\sqrt{3}}}-
k^2{{\left( 3 + 4{{C_3}^2} \right) }\over {9{\sqrt{3}}C_3}}-
k^4{{\left( 3 + 16\,{{C_3}^4} \right) }\over
   {324{\sqrt{3}}{{C_3}^3}}}\cr
-& k^6{{27 + 108{{C_3}^2} - 72{{C_3}^4} +800{{C_3}^6}}\over
   {52488{\sqrt{3}}{C_3}^5}}+O(k^8)\cr
\noalign{\smallskip}
e^{(4)}=&-1-2C_4-
k^2\left({1\over 4} + {{C_4}\over 8} +
 {1\over {2\left( 1 + C_4 \right) }}\right)\cr
-&k^4\left({1\over {64}} + {{9C_4}\over {512}} +
  {1\over {32{{\left( 1 + C_4 \right) }^3}}} +
  {3\over {64{{\left( 1 + C_4 \right) }^2}}}\right)+O(k^6)\cr
\noalign{\smallskip}
e^{(5)}=&-(5- \sqrt{5})C_5^3 +(7- \sqrt{5}) C_5\cr
 +&k^2\biggl({1\over {5\,\left( -5 + 2\,{\sqrt{5}} \right)
      \left( -1 + C_5 \right) }} +
  {{2\,\left( -3 + {\sqrt{5}} \right) }\over
    {5\,\left( -5 + 2\,{\sqrt{5}} \right) C_5}} +
  {1\over {5\,\left( -5 + 2\,{\sqrt{5}} \right)\left( 1 + C_5 \right)}}\cr
-&{{C_5\,\left( 18 - 8\,{\sqrt{5}} - 7\,{{C_5}^2} +
3{\sqrt{5}}\,{{C_5}^2} \right) }\over
 {5\left( -5 + 2{\sqrt{5}} \right) }}\biggr)+O(k^4) \cr
\noalign{\smallskip}
e^{(6)}=&3-{4\over 3}C_6-{8\over 3}C_6^2-k^2 \bigl({2\over 27}C_6^2+{4\over
27}C_6+5/36+{1\over C_6}-{4\over 3(3+2C_6)} \bigr)+O(k^4)}$$

\vfill
\eject
{\bf Table~6}
The first six orders of perturbation theory in $k$ for $N=3$ for the
correlation function $\langle Z_0 Z_r^{\dagger}\rangle=\sum_{n=0} k^n
c^z_n$ in the generic case. We use the notation $C_3=\cos\phi/3$ and
$S_3=\sin\phi/3.$ For $c^z_5$ only the coefficient of $\delta_{r,1}$
and for $c^z_6$ only the coefficient of $\cos^2{\bar
\phi}\delta_{r,1}$ is given.

$$\ninepoint\eqalign{
c^z_0&=1,~~~c^z_1=0,~~~
c^z_2={1\over 6 C_3^2}(\delta_{r,0}-1),~~~
c^z_3={\cos {\bar \phi}\over 18 C_3^3}(\delta_{r,0}-1)\cr
\noalign{\smallskip}
c^z_4&=\bigl({1\over 432 C_3^4}+{2\over 81 C_3^2}-{2\over 27(1-2C_3)^2}-
{2\over 81(1-2C_3)}-{2\over 27(2C_3+1)^2}-{2\over 81
(2C_3+1)}\bigr)(1-\delta_{r,0})\cr
&\bigl({5\over 432 C_3^4}+{1\over 81 C_3^2}+{2i S_3\over 27
C_3}\cr
&+{1-2iS_3\over 54(2C_3+1)^2}+{7-12iS_3\over
162(2C_3+1)}+{1+2iS_3\over 54 (2C_3-1)^2}+{7+12iS_3\over
162(2C_3-1)}\bigr)\delta_{r,1}\cr
\noalign{\smallskip}
c^z_5&=\cos{\bar \phi}\bigl({1\over 648 C_3^5}-{1\over 81C_3^3}+{2\over
81 C_3}+{2iS_3\over 27 C^2_3}\cr
&+{1+2iS_3\over 27(2C_3-1)^2}-{2+18iS_3\over 81
(2C_3-1)}+{-1+2iS_3\over 27(2C_3+1)^2}+{-2+18iS_3\over
2C_3+1}\bigr)\delta_{r,1}+\cdots\cr
\noalign{\smallskip}
c^z_6&=\cos^2{\bar \phi}\bigl({-1\over 1296 C_3^6}-{1\over 108
C_3^4}-{1\over 54 C_3^2}+{iS_3\over 54 C_3^3}+{4i S_3\over 27 C_3}\cr
&+{1-2iS_3\over 54 (2C_3+1)^2}-{1+8i S_3\over54(2C_3+1)}+{1+2i S_3\over
54(2C_3-1)^2}+{1-8i S_3\over 54 (2C_3-1)}\bigr)\delta_{r,1}+\cdots}$$

\vfill

\eject

{\bf Table~7}
The first five sums, $S_{2,n}(N)$, from~\trigidh~with $P_{2,n}(N,j,k)$
given by~\trigidc.

$$\ninepoint\eqalign{
S_{2,0}(N) & = -N^2 (N^2-1) (N^2-4) (53 N^2+13) /
(16 \cdot 27 \cdot 5 \cdot 7) \cr
S_{2,1}(N) & = N^2 (N^2-1) (N^2-4) (821 N^4+280 N^2+69) / (8 \cdot 81 \cdot
25 \cdot 7) \cr
S_{2,2}(N) & = -N^2 (N^2-1) (N^2-4) (30874 N^6+11150 N^4+3501 N^2+845) /
(32 \cdot 81 \cdot 5 \cdot 7 \cdot 11) \cr
S_{2,3}(N) & = N^2 (N^2-1) (N^2-4) (103801414 N^8 + 37967420 N^6 +
12484059 N^4 + 3718690 N^2 + \cr
 & 882407) / (8 \cdot 27 \cdot 125 \cdot 7 \cdot 11 \cdot 13
 \cdot 17) \cr
S_{2,4}(N) & = -N^2 (N^2-1) (N^2-4) (1012869789 N^{10} + 371602425 N^8 +
       123446295 N^6 + \cr
 & 38256875 N^4 + 11003386 N^2 + 2579010) /
(32 \cdot 81 \cdot 5 \cdot 7 \cdot 11 \cdot 13 \cdot 31)}$$

\vfill

\eject

{\bf Table~8}
The first three members of the third family and the first two members
of the fourth family of solutions to~\trigidh~expressed in terms of
the Bernoulli polynomials and using the notation $J=j/N$, $K=k/N$.

$$\ninepoint\eqalign{
P_{3,0}(N,j,k) & = N^9 (J^4 (K-2 K^3+K^5) - 2 J^5 (K-K^3) +
 J^6 (K+1/3 K^3)) \cr
P_{3,1}(N,j,k) & = N^{11} (1/4 J^4 (K-6 K^3+9 K^5-4 K^7) +
 3/2 J^5 (K^3-K^5) - \cr
 &         1/4 J^6 (6 K-9 K^3+7 K^5) + 2 J^7 (K-K^3) -
 1/4 J^8 (3 K+K^3)) \cr
P_{3,2}(N,j,k) & = N^{13} (J^4 (5/9 B_9(K)+B_7(K)) +
J^5 (1/12 B_7(K)-1/36 B_3(K)) + \cr
 & J^6 (2 B_7(K)+41/24 B_5(K)-1/8 B_3(K)) + J^7 3/4 B_5(K) + \cr
 & J^8 (4/3 B_5(K)+41/72 B_3(K)-1/16 B_1(K)) +
J^9 5/9 B_3(K) + \cr
 & J^{10} (1/9 B_3(K)+1/9 B_1(K))) \cr
P_{4,0}(N,j,k) & = N^{12} (J^5 (K-3 K^3+3 K^5-K^7) -
3 J^6 (K-2 K^3+K^5) + \cr
 &         J^7 (3 K-2 K^3-K^5) - J^8 (K+K^3)) \cr
P_{4,1}(N,j,k) & = N^{14} (J^5 (5/9 B_9(K)+B_7(K)) +
J^6 (2/3 B_7(K)+1/2 B_5(K)-1/18 B_3(K)) + \cr
 & J^7 (14/9 B_7(K)+17/9 B_5(K)) +
J^8 (4/3 B_5(K)+1/2 B_3(K)-1/36 B_1(K)) + \cr
 & J^9 (5/9 B_5(K)+20/27 B_3(K)) + J^{10} (2/9 B_3(K)+1/9 B_1(K)))}$$

\vfill

\eject

{\bf Table~9}
The sums~\trigidh~corresponding to the polynomials in Table~8.

$$\ninepoint\eqalign{
S_{3,0}(N) & = N (N^2-1) (N^2-4) (N^6+5N^4+54N^2-80) \cdot 32 /
(27 \cdot 25 \cdot 7 \cdot 11) \cr
S_{3,1}(N) & = N (N^2-1) (N^2-4) (9N^8+45N^6+46N^4+12920N^2- \cr
 & 31680) \cdot 8 / (27 \cdot 25 \cdot 7 \cdot 11 \cdot 13) \cr
S_{3,2}(N) & = -N (N^2-1) (N^2-4) (198N^{10}+990N^8+7980N^6+ \cr
 & 58105N^4+15362N^2-786240) / (64 \cdot 243 \cdot 25 \cdot 49 \cdot
11 \cdot 13) \cr
S_{4,0}(N) & = N^2 (N^2-1) (N^2-4) (2N^8+10N^6+42N^4+2315N^2- \cr
 & 9614) \cdot 256 / (27 \cdot 25 \cdot 49 \cdot 11 \cdot 13) \cr
S_{4,1}(N) & = -N^2 (N^2-1) (N^2-4) (9N^{10}+45N^8+189N^6+ \cr
 & 7915N^4+11792N^2-240240) / (32 \cdot 243 \cdot 25 \cdot 49 \cdot
 11 \cdot 13)}$$

\vfill

\eject

\listrefs

\vfill\eject

\bye
\end